\documentclass[journal]{IEEEtran}
\usepackage{amsmath,amsfonts}
\usepackage{algorithmic}
\usepackage{array}
\usepackage[caption=false,font=normalsize,labelfont=sf,textfont=sf]{subfig}
\usepackage{textcomp}
\usepackage{stfloats}
\usepackage{url}
\usepackage{verbatim}
\usepackage{graphicx}
\usepackage{cite}
\def\BibTeX{{\rm B\kern-.05em{\sc i\kern-.025em b}\kern-.08em
    T\kern-.1667em\lower.7ex\hbox{E}\kern-.125emX}}
\usepackage{balance}
\graphicspath{{figures/}}

\begin{document}
\title{Self-calibrated Microring Weight Function for Neuromorphic Optical Computing}
\author{Jos\'e Garc\'ia Echeverr\'ia*, \textit{Graduate Student Member, IEEE}, Daniel Musat*, Ataollah Mahsafar, Kaveh R. Mojaver, \textit{Member, IEEE}, David Rolston, \textit{Member, IEEE}, Glenn Cowan \textit{Member, IEEE}, Odile Liboiron-Ladouceur \textit{Senior Member, IEEE}

\thanks{This work was funded by Axonal Networks and the MITACS Accelerate program (IT27043), and subsidized by CMC Microsystems which provided CAD tools and fabrication services using the Silicon Photonics technology platform from Advanced Micro Foundry (AMF). (\textit{Equally contributing authors: José García Echeverría and Daniel Musat}).

José García Echeverría is with the Electrical and Computer Engineering Department of McGill University, Montreal, QC Canada (e-mail: jose.garciaecheverria@mail.mcgill.ca). 

Daniel Musat is with the Electrical and Computer Engineering Department of Concordia University, Montreal, QC Canada (email: daniel.musat@mail.concordia.ca).

Color versions of one or more of the figures in this article are available online at http://ieeexplore.ieee.org}}

\markboth{Journal of Lightwave Technology Class Files,~Vol.~XX, No.~XX, XXXX~2024}%
{How to Use the IEEEtran \LaTeX \ Templates}

\maketitle

\begin{abstract}
This paper presents a microring resonator-based weight function for neuromorphic photonic applications achieving a record-high precision of 11.3 bits and accuracy of 9.3 bits for 2 Gbps input optical signals. The system employs an all-analog self-referenced proportional-integral-derivative (PID) controller to perform real-time temperature stabilization within a range of up to 60 ºC. A self-calibrated weight function is demonstrated for a range of 6 ºC with a single initial calibration and minimal accuracy and precision degradation. By monitoring the through and drop ports of the microring with variable gain transimpedance amplifiers, accurate and precise weight adjustment is achieved, ensuring optimal performance and reliability. These findings underscore the system's robustness to dynamic thermal environments, highlighting the potential for high-speed reconfigurable analog photonic networks.
\end{abstract}


\section{Introduction}
\IEEEPARstart{I}{n} the domain of neuromorphic computing, researchers strive to develop architectures that replicate the remarkable abilities of the human brain \cite{mead}. However,  traditional computing models, such as the Von Neumann architecture, face challenges in energy efficiency, fault tolerance, and parallel processing \cite{hamilton,marr,merolla}.This has sparked interest in photonics, which offers inherent advantages like low latency and power dissipation, making it a promising avenue for neuromorphic systems across various fields, including machine learning \cite{shen}, non-linear programming \cite{han}, and signal processing \cite{trait}. 

Converting neural network models into photonic platforms presents unique hurdles, particularly in creating tunable and compact synaptic weights crucial for scalable learning and cognitive functions within photonic neuromorphic networks. Microring resonators (MRRs) have emerged as promising solutions due to their wide tuning range, reconfigurability, and small size \cite{xiao,xu}. However, their spectral properties are closely tied to their refractive effective index, making them sensitive to fabrication variations \cite{lu} and environmental factors, notably temperature changes \cite{lipson}. In particular, these variations have a direct impact on the refractive index of the waveguide structure. It is essential to note that MRRs exhibit an exceptionally high sensitivity to alterations in the refractive index of their constituent waveguides. Minute deviations in this index lead to shifts in the resonance peak of the microring. In some cases, these shifts may amount to several hundreds of picometers or, even to the nanometer range \cite{poon}. Such fluctuations are undesirable, underlining the need for wavelength tuning and real-time stabilization of microring-based devices.

Typically, microring controllers are engineered to operate the devices as optical filters or modulators, with stabilization points targeting the maximum extinction ratio (ER), or optical modulation amplitude (OMA), or minimize transmitter penalty (TP) using a variety of stabilization techniques for their desired functionality. Some of these circuits rely on low responsivity in-resonator defect state absorption (DSA). In this technique the light circulating inside a microring’s waveguide interacts with defects within the crystal structure created during the doping processes. The generated photocurrent is used to monitor the amount of light trapped inside the microring \cite{murray}. Other techniques rely on maximum/minimum point searching \cite{watts,zheng}, homodyne locking \cite{cox}, dithering signals \cite{padmaraju}, high-speed bit-rate electronics leading to high power dissipation \cite{zortman} or requiring non-trivial changes in the fabrication process \cite{guha}. 

In applications such as optical analog computing, microring control is required to dynamically reprogram the stabilization point of the device. This adaptability is essential to accommodate learning functions within the system, allowing for real-time adjustments to optimize performance and support evolving computational tasks. The dynamic tunability of the optical resonance, along with the requirement for a high number of microrings to implement synaptic connections in neuromorphic photonics, poses significant challenges to the microring control which is particularly evident during the initial training and calibration processes \cite{zortman,prucnal,zhou}. The choice of an appropriate tuning, stabilization and control circuits profoundly influences the overall performance and scalability of a microring based system.

This research focuses on establishing an all-analog, self-referenced, and automatic electronic feedback circuit for microring stabilization with a single microring and one wavelength per waveguide. Its functions include microring’s resonance tuning to the operating laser wavelength, real-time temperature monitoring and stabilization, and dynamic control of the optical spectrum for neuromorphic computing applications. The designed circuit achieves self-calibrated weights for changes in temperature for a 6 ºC range maintaining accuracy and precision resolution slightly above 9 bits and 11 bits, respectively. The paper is structured as follows: section II presents the concept of a weight function in neural networks. We delve into the design of the MRR, and the thermal tuning mechanism employed. Additionally, we describe the control strategy devised for regulating the weight function and the calibration method. Section III presents our experimental results for the thermal tuning and the weight function measurements. In Section IV, we analyze the experimental outcomes and propose potential future enhancements. Finally, Section V summarizes the key results of our work.

\section{Weight Function and Controller Design}
\subsection{Weight Function}
\begin{figure}[!ht]
\centering
\includegraphics[width=3.4in]{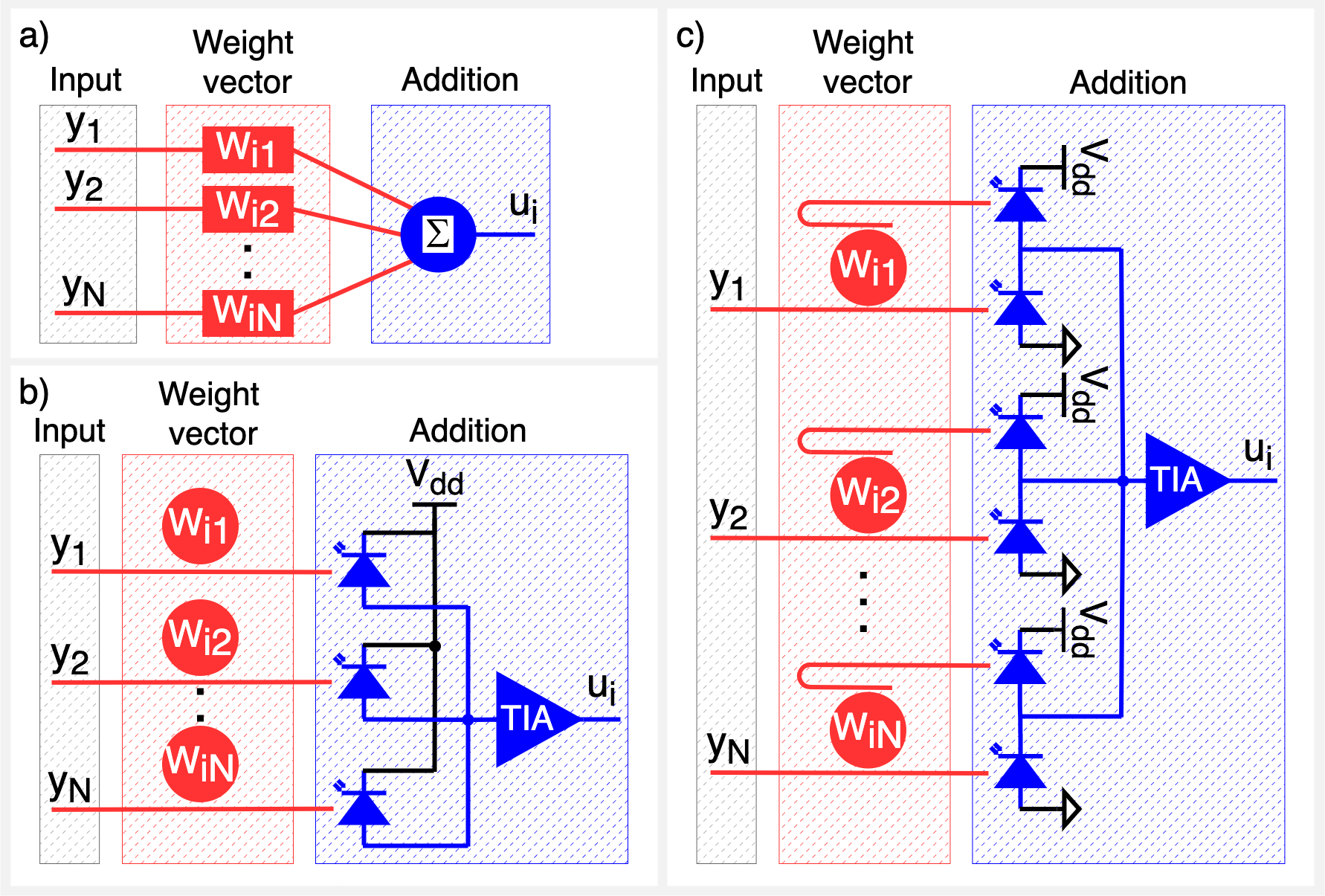}
\caption{Weighted addition function with a single microring and one wavelength per waveguide represented as a) a generic building block of a weighted addition function, b) a single-ended microring-based weighted addition function, and c) a differential microring based weighted addition function.}
\label{fig1}
\end{figure}

\noindent Neural networks consist of interconnected nonlinear elements, commonly referred to as neurons, with adjustable linear weights known as synapses \cite{mead}. The synapses are implemented with tunable weights  \cite{hamilton,merolla}, allowing for dynamic adaptation and learning within neuromorphic photonic systems. Specifically, a weighted addition function, illustrated in Fig. 1a, can be implemented using a bank of dynamically spectrally tuned MRRs to enable learning and plasticity. By tuning to the desired wavelength and adjusting the microring’s intracavity phase shifts, localized circuits can dynamically modify the strength of synaptic connections, or weights, to enable adaptability and neural network functionality  \cite{mead,hamilton,mitra}. The fundamental expression for the weighted addition function is expressed as follows:

\begin{equation}
\label{eq1}
u_i = \vec{w_i} \cdot \vec{y_i} \quad \forall \; i \;\in\; [1,N]
\end{equation}

where $u_i$ is the output light intensity, $\vec{w_i}$ is the weight vector, $\vec{y_i}$ is the input light intensity and $N$ is the weight vector size. In an MRR-based system, the weight vector $\vec{w_i}$ can be realized in either single-ended or differential configurations. In the single-ended setup (Fig. \ref{fig1}b), the intensity at either the through or drop port yield weight values between 0 and 1. Conversely, the differential configuration (Fig. \ref{fig1}c) employs balanced photodetectors resulting in weight values spanning from -1 to 1.

\subsection{Microring Resonator Design}
\begin{figure}[!ht]
\centering
\includegraphics[width=3.4in]{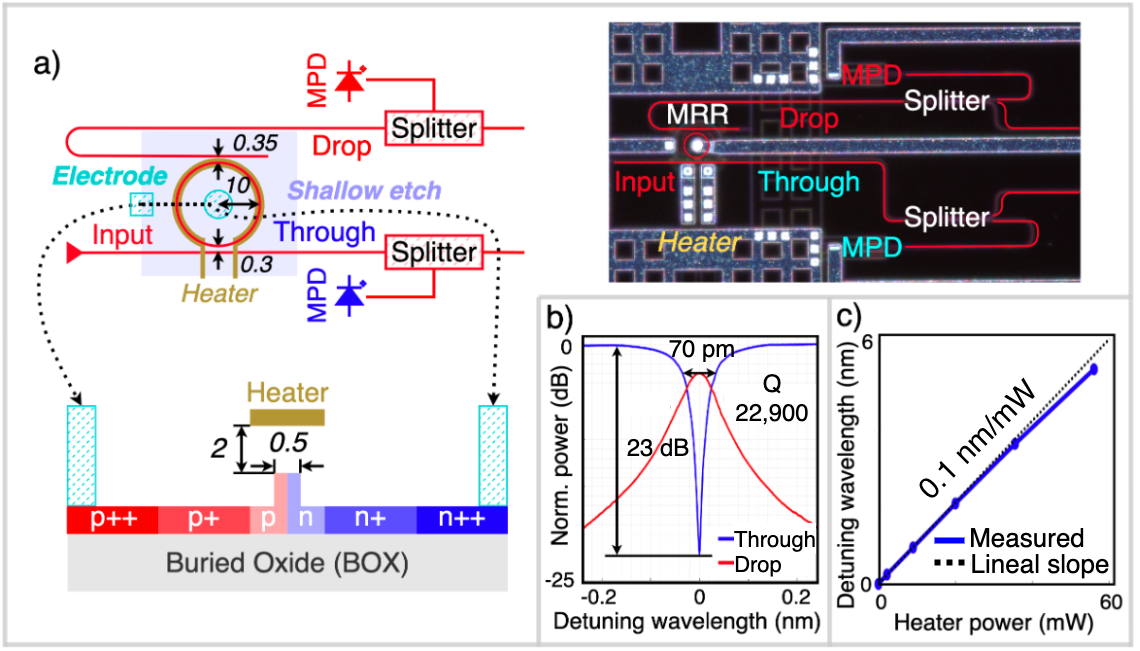}
\caption{a) Schematic, cross-section, and micrograph of the photonic circuit with double bus MRR and through/drop monitor photodetectors (MPD) to implement tuning, stabilization and weight function with thermal feedback circuit, b) measured transmission spectrum of the MRR, and c) measured thermal shift efficiency.}
\label{fig2}
\end{figure}

\noindent This work requires a double bus MRR with power taps at both through and drop ports that will serve as inputs for the thermal feedback stabilization circuit (TFSC) to tune and stabilize the microring. The double bus MRR is designed based on \cite{guan} and fabricated in an AMF silicon-on-insulator (SOI) 193 nm lithography technology platform with a silicon layer thickness of 220 nm and buried oxide thickness of 3 $\mu$m. The MRR waveguide is etched to a 90 nm thick slab and has 500 nm width for single-mode operation, an insertion loss of approximately 2 dB, a radius of 10 $\mu$m, and a measured free spectral range (FSR) of 8.8 nm. The schematic representation of the designed MRR with a cross-section view and a micrograph are shown in Fig. \ref{fig2}a. 

Critical coupling maximizes extinction ratio leading to a higher dynamic range for the microring weight function. Due to process variations, the microring is designed slightly under coupled with through and drop gaps of 300 nm and 350 nm, respectively, and includes coupling regime correction electrodes connected to the microring’s waveguide. Applying reverse bias to the MRR waveguide using the coupling correction electrodes reduces the propagation losses inside the ring, thus moving from under coupling to critical coupling regime. The fabricated MRR features a measured Q-factor of 22,900, a full-width half-maximum (FWHM) of 70 pm, and an extinction ratio (ER) of 23 dB as illustrated in Fig. \ref{fig2}b. For the thermal tuning, a titanium-tungsten (TiW) metal layer is included 2 $\mu$m above 90\% of the microring’s waveguide circumference with a nominal resistance of 450 $\Omega$, resulting in a thermal tuning efficiency of approximately 100 pm/mW as shown in the thermooptic characterization results (Fig. \ref{fig2}c). The thermal tuning efficiency could be improved with isolation trenches \cite{dong} that confine the heat within the microring structure. Additionally, waveguide semiconductor heaters \cite{luo,harris} could enhance the efficiency by achieving the thermooptic phase shift through the heat generated due to resistive losses when an electrical current is applied through the microring semiconductor waveguide. These methods result in tuning efficiencies above 1 nm/mW. 

\subsection{Thermal Tuning Operation}
\begin{figure}[!ht]
\centering
\includegraphics[width=3.4in]{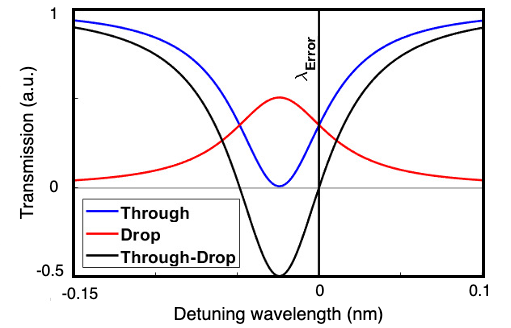}
\caption{Optical transmission spectrum versus detuning wavelength of a double bus MRR with through, drop and the through minus drop (error signal) curves. The wavelength where the error signal equals to zero is represented by $\lambda_{error}$ and determines the zero value for detuning wavelength axis.}
\label{fig3}
\end{figure}

\noindent To explain how the proposed thermal stabilization circuit works, Fig. \ref{fig3} illustrates the transmission spectrum of both outputs of a double-bus MRR with a continuous-wave (CW) laser at the input. The transmission curves of through (blue) and drop (red) follow the equations described by Eq. \ref{eq2} and Eq. \ref{eq3} \cite{bogaerts}:

\begin{equation}
\label{eq2}
T_T = \frac{r_2^2 a^2 - 2 r_1 r_2 a \cos{\phi}+r_1^2}{1-2 r_1 r_2 a \cos{\phi} + (r_1 r_2 a)^2}
\end{equation}

\begin{equation}
\label{eq3}
T_D = \frac{k_1^2 k_2^2a}{1-2 r_1 r_2 a \cos{\phi} + (r_1 r_2 a)^2}
\end{equation}

where $T_T$ is the transmitted intensity of through, $T_D$ is the transmitted intensity of drop, $\phi=\beta L$ is the single-pass phase shift, with $L$ the round-trip length and $\beta$ the propagation constant of the circulating mode. $r_1$ and $ r_2$ are the self-coupling coefficients of through and drop, respectively, $k_1$ and  $k_2$ are the cross-coupling coefficients of through and drop, respectively, and $a$ is the single-pass amplitude transmission. 

The input signal to the TFSC is the difference between the through power and the drop power, represented by the solid black line in Fig. \ref{fig3}. The error signal is symmetric with respect to the resonance notch of the microring spectrum, with each side being a monotonic signal that crosses zero. This property allows for a feedback controller that varies the power applied to the integrated heater to lock the operation of the MRR at the wavelength ($\lambda_{Error}$) where the error signal is zero or, in other words, where the optical power at the through and drop outputs are equal. Waveguide taps sample the output powers used for the tuning circuit.

\begin{figure}[!ht]
\centering
\includegraphics[width=3.4in]{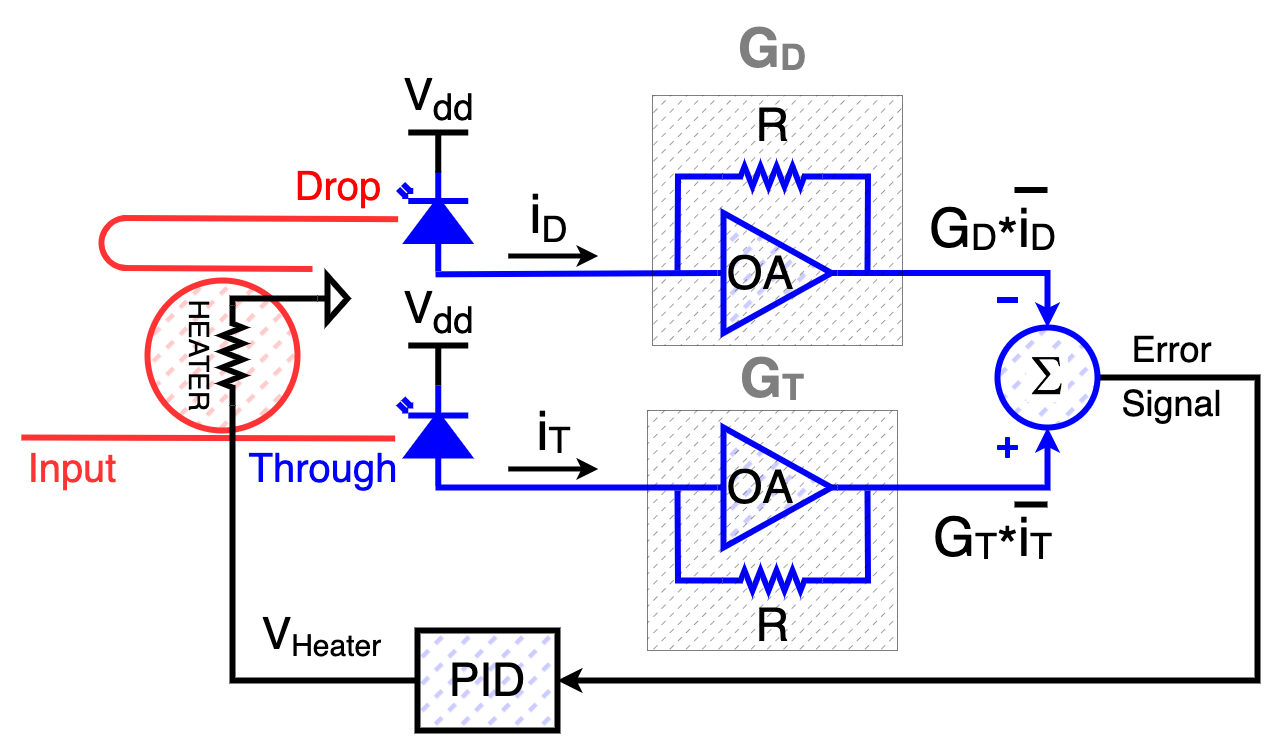}
\caption{Block diagram of the thermal feedback stabilization circuit using through and drop monitoring photodetectors, transimpedance amplifiers, subtractor and proportional-integral-derivative controller that drives the integrated heater of the MRR.}
\label{fig4}
\end{figure}

The TFSC shown in Fig. \ref{fig4} requires a double bus MRR with monitoring photodetectors (MPDs) for the through and drop ports. The MPDs monitor the optical power at the outputs of the microring and the generated photocurrents provide the input signals for the self-referenced thermal feedback tuning, stabilization, and control circuit. The resulting photocurrents are amplified by low speed transimpedance amplifiers (TIAs). Subsequently, the difference between these signals serves as an error signal to feed a PID controller. The output of the PID is then used to vary the power supplied to the integrated TiW metal heater for temperature changes within the waveguide of the microring structure. As a consequence of this temperature alteration, the refractive index of the microring’s waveguide is modified leading to a shift of the resonant peak wavelength given the thermooptic effect in silicon \cite{cocorullo,dwivedi}. Finally, the microring’s resonant peak shifts until the wavelength of the CW laser at the input of the microring matches the point where the difference between the photocurrents becomes zero, resulting in a precise and dynamic adjustment of the microring’s resonant frequency. 
The default operating point of the MRR occurs for TIA through gain ($G_T$) equal to TIA drop gain ($G_D$). This condition is set to achieve automatically a balance where the difference between the amplified photocurrents from the through and drop ports is minimized. This effectively leads to an equilibrium point where the average transmitted intensity of through ($T_T$) equals the average transmitted intensity of drop ($T_D$) multiplied by the respective TIA gains and photodetector responsivities $R_T$ for through and $R_D$ for drop as expressed in Eq. \ref{eq4}.

\begin{equation}
\label{eq4}
G_T R_T [r_2^2 a^2 - 2 r_1 r_2 a \cos{\phi}+r_1^2] = G_D R_D [k_1^2 k_2^2a]
\end{equation} 

\begin{figure}[!ht]
\centering
\includegraphics[width=3.4in]{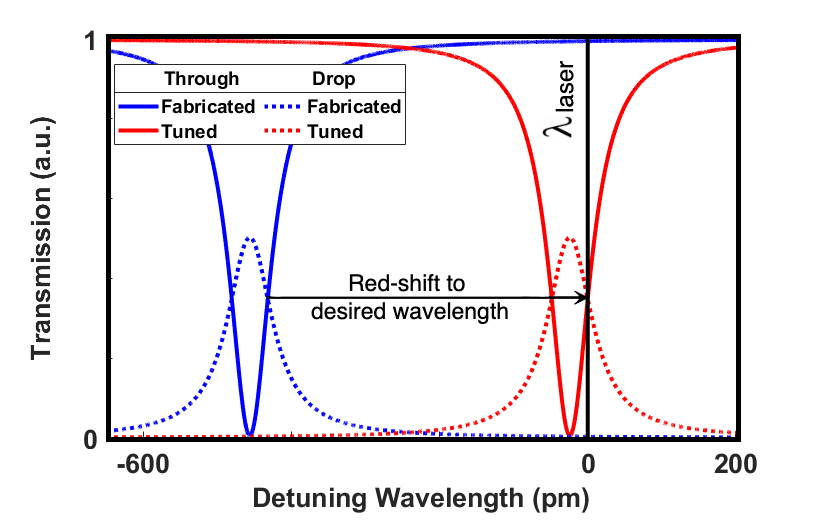}
\caption{Redshift of a fabricated MRR optical spectrum with power applied to the integrated heater to achieve laser wavelength tuning at equal through and drop intensities.}
\label{fig5}
\end{figure}

For the specific case of $G_T R_T= G_D R_D$, the locking point is achieved at the location of the optical spectrum where the optical power of the through port and the drop port are equal. Fig \ref{fig5} shows the MRR optical spectrum before (blue) and after (red) the TFSC locks the MRR. By applying power to the integrated heater, the MRR optical spectrum is red shifted to the desired operating point represented by Eq. \ref{eq4} corresponding to the crossing of the through and drop spectrums at the locking laser wavelength.

\subsection{Weight Function Control}
\begin{figure}[!ht]
\centering
\includegraphics[width=3.4in]{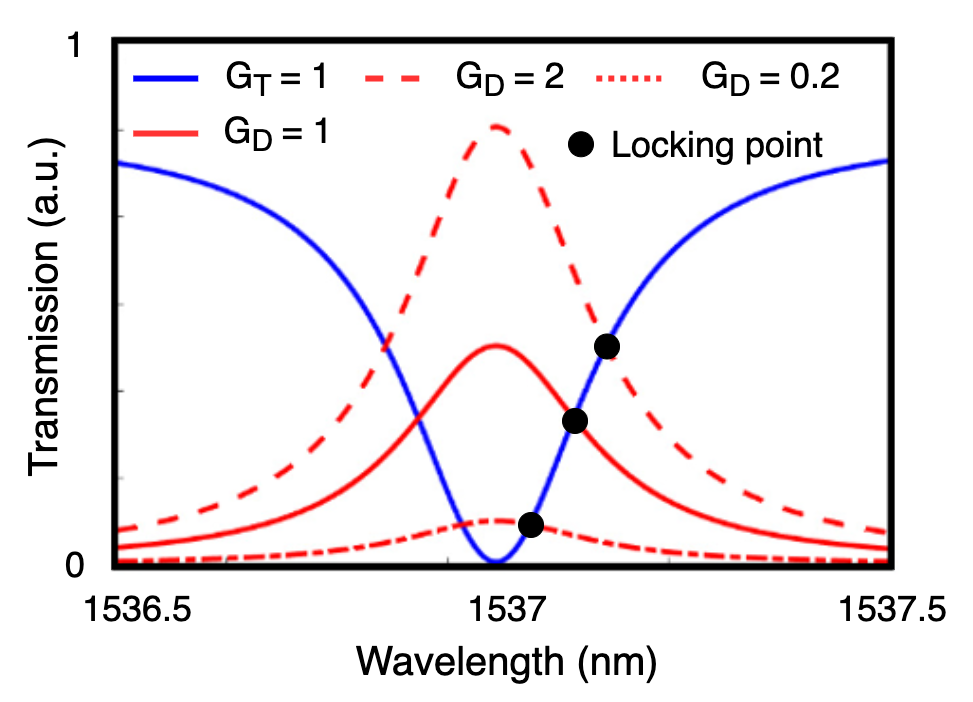}
\caption{Graphical representation of the drop transmission when the drop TIA gain is varied while keeping through TIA gain constant.}
\label{fig6}
\end{figure}

\noindent In principle, it is possible to precisely target transmission values within the spectrum of the MRR by adjusting the relationship between the gains of the through ($G_T$) and drop ($G_D$) TIAs. By maintaining a constant through gain and adjusting only the drop gain, the crossing point of the spectrum shifts, enabling precise locking at any desired location on the transmission spectrum. This is illustrated in Fig. \ref{fig6}, where the drop port gain is varied, and the crossing point, i.e., locking point, represented as a black dot changes.

\begin{figure}[!ht]
\centering
\includegraphics[width=3.4in]{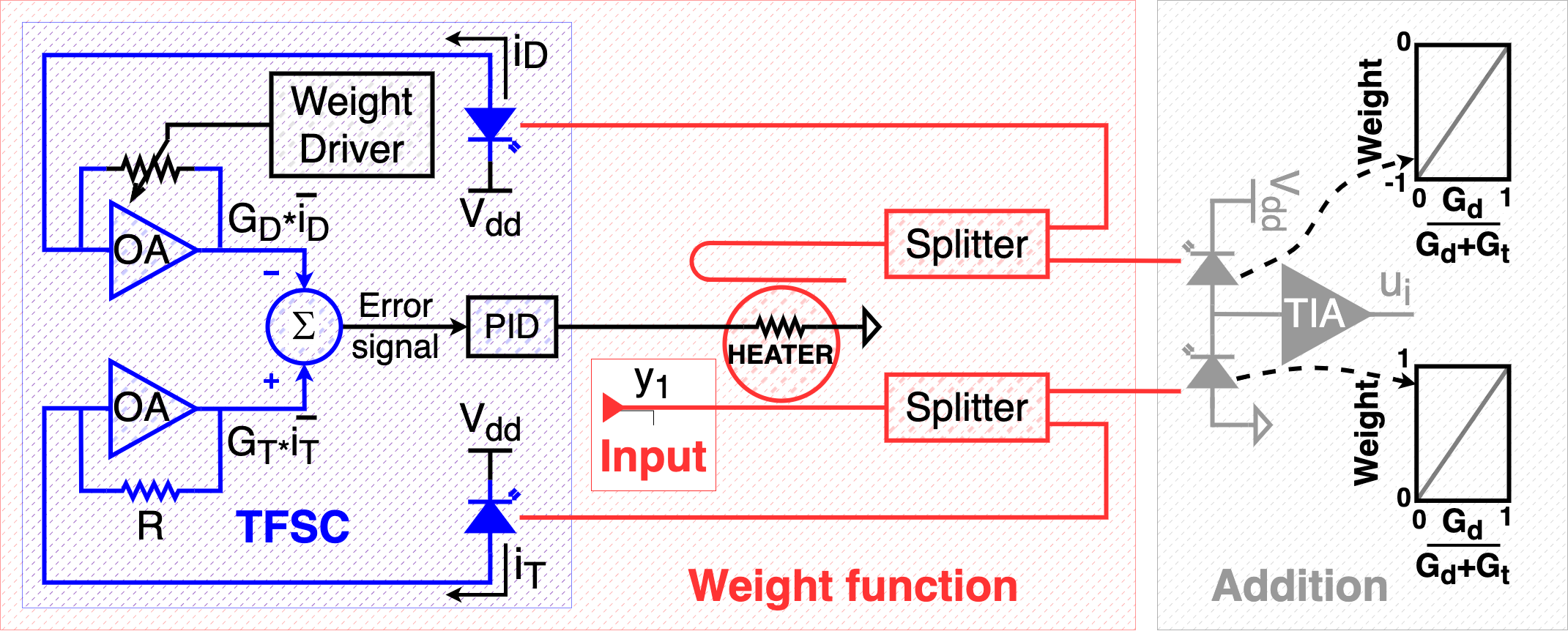}
\caption{Thermal tuning, real-time stabilization, and weight control circuit to implement a differential self-calibrated weight function.}
\label{fig7}
\end{figure}

This gain relationship can be adjusted by changing the feedback resistor of the low-speed TIAs using a local circuit, i.e., weight driver. In our solution, this driver provides a preset value from the neural network to vary the feedback resistance of the drop TIA, as depicted in Fig. \ref{fig7}. Given the fact, that our system is made with off-the-shelf components mount on a PCB, the manufactured solution does not give an objective view of the achievable performance. The scope of the paper is focused on a proof-of-concept demonstration, specifically, our design proves that the novel self-calibrated control combined with a photonic integrated circuit with MRRs is suitable for neuromorphic applications.

\subsection{Calibration Method}
\noindent Due to manufacturing process variations, the simulated optical transmission does not match the measured transmission of the fabricated device, leading to inaccuracies in the weight mapping function. Therefore, an initial calibration is needed after the MRR device is manufactured. In particular, previous designs employing in-resonator photoconductive heaters (IRPH) require calibration based on multiple variables including input optical power, electrical crosstalk, and thermal crosstalk \cite{huang}. While such considerations proved effective in partially alleviating accuracy resolution loss, they often required exhaustive initial characterization and calibration efforts \cite{lima}. In this work, the initial calibration is significantly simplified due to the nature of the thermal feedback circuit with a self-referenced PID controller. After the initial calibration is performed, the designed solution features self-calibration to adapt to real-time environmental thermal perturbations, input optical power changes, and electrical crosstalk caused by common ground parasitic resistance.

\begin{figure}[!ht]
\centering
\includegraphics[width=3.4in]{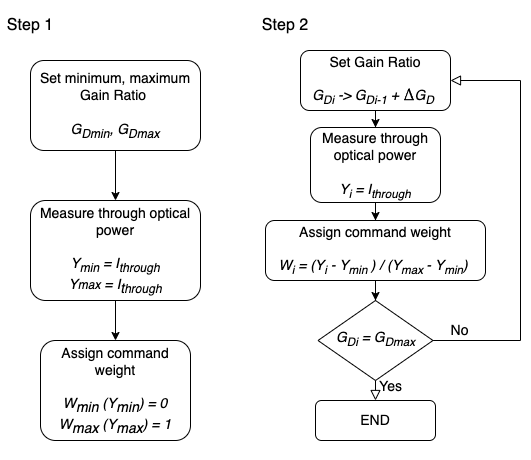}
\caption{Flow diagram of the two-step calibration algorithm. Step 1 consisting of determining the dynamic range of the weight function $W_{min}$ and $W_{max}$ by applying the minimum and maximum drop transimpedance amplifier gains $G_{Dmin}$ and $G_{Dmax}$, respectively. In Step 2, the command weights are assigned by sweeping the drop transimpedance amplifier gain $G_{Di}$ with $\delta G_D$ gain steps and normalizing based on the dynamic range found on step 1 until reaching $G_{Dmax}$.}
\label{fig8}
\end{figure}

Our initial two-step calibration is reported in the flow diagram of Fig. \ref{fig8}. The first step determines the dynamic range of the weight function. The minimum gain achievable by the weight driver ($G_{Dmin}$) translates into an optical output intensity and corresponds to the weight value of zero (0). Similarly, the maximum gain ($G_{Dmax}$) corresponds to the weight value of one (1). The dynamic range is expressed as the difference between the maximum and the minimum transmitted optical powers that are functions of the maximum and minimum gains.

\begin{equation}
\label{eq5}
DR = I_{max}(G_{Dmax})-I_{min}(G_{Dmin})
\end{equation}

In the second step, the gain is swept from $G_{Dmin}$ to $G_{Dmax}$ and the transmitted optical power at each gain step $\Delta G_D$ of the sweep are measured. A command weight is assigned and registered to each measured intensity. The resulting command weight is based on the dynamic range of the optical transmissions achievable with the variable resistors in the TIA path derived from the expression:

\begin{equation}
\label{eq6}
W_i = \dfrac{I(G_{Di})-I_{min}(G_{Dmin})}{DR}
\end{equation}

Where $W_i$ is the weight value of the i-th step of the sweep and $I(G_{Di})$ is the transmitted optical intensity for the i-th step drop gain of the sweep. When the sweep reaches the maximum gain relationship ($G_{Dmax}$) the initial calibration finalizes. 

During real-time operation, the circuit senses the perturbations caused by either electrical and thermal crosstalk that vary the through and drop intensity relationship. The closed-loop feedback automatically compensates for both these perturbations by continuously adjusting the integrated heater power to match the monitored through and drop intensities of the microring with the pre-recorded values obtained during the previously described initial two-step calibration process. Similarly, the initial calibration allows the system to operate under conditions of an asymmetric Lorentz spectrum or limited extinction ratio, as the recorded values during calibration already account for these effects.

\section{Experimental Results}
\begin{figure}[!ht]
\centering
\includegraphics[width=3.4in]{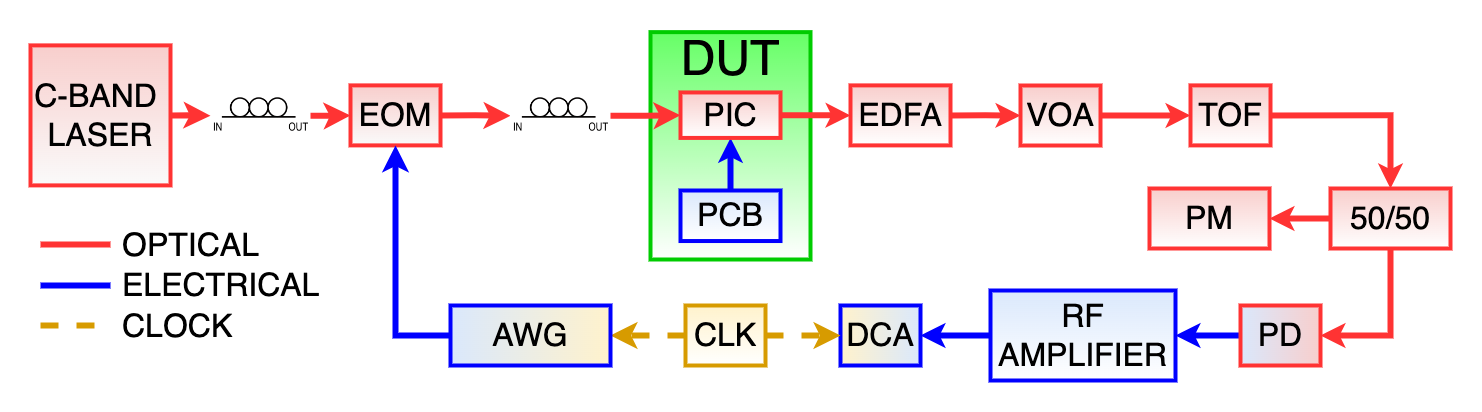}
\caption{Experimental setup to perform the weight function measurements including continuous-wave C-band laser, electrooptic modulator (EOM), arbitrary waveform generator (AWG) with integrated clock signal (CLK), device under test (DUT) including the photonic integrated circuit (PIC) and a printed circuit board (PCB), erbium-doped fiber amplifier (EDFA), variable optical attenuator (VOA), tunable optical filter (TOF), 50/50 power splitter, power meter (PM), high-speed photodetector (PD), RF amplifier and a digital communication analyzer (DCA).}
\label{fig9}
\end{figure}

\noindent The experimental setup shown in Fig. \ref{fig9}, comprises a C-band laser that generates the optical carrier, which is subsequently modulated by a 2 Gbps PRBS31 signal sourced from a Keysight Arbitrary Waveform Generator (AWG) M8199A. The optical modulation is achieved using a Powerbit F10 Oclaro electrooptic modulator (EOM). The light is coupled in and out of the device under test (DUT) via vertical grating couplers, leading to a total loss of 14 dB. The EOM and the grating couplers are polarization sensitive, so polarization controllers (PC) are included before and after the EOM. An erbium-doped fiber amplifier (EDFA) is used at the output of the PIC to compensate for the coupling loss and insertion loss of the EOM. To prevent high optical powers into the power meter (PM) and photodetector (PD) and reduce the amplified spontaneous emission (ASE) noise generated by the EDFA, a variable optical attenuator (VOA) and a tunable optical filter (TOF) are included after the EDFA. The resulting optical signal is measured using a PM for the average power and a 50 $\Omega$ terminated Finisar photodetector (PD) XPDV2320R connected to an RF amplifier SHF 826H followed by a Keysight 86112A-HBW DCA for the signal dynamics. The DUT consisted of the PIC containing passive and active optical devices and the thermal feedback stabilization circuit that is built on a printed circuit board (PCB) and wirebonded to the PIC.

\subsection{Thermal Tuning}
\begin{figure}[!ht]
\centering
\includegraphics[width=3.4in]{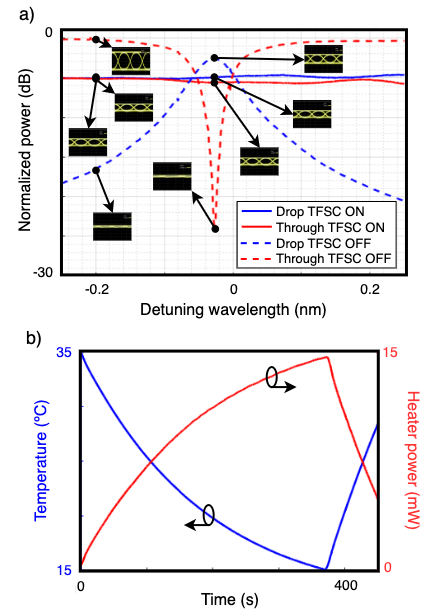}
\caption{a) Measured optical spectrum of a microring modulator after a laser sweep w/ TFSC ON and w/ TFSC OFF, and b) temperature sweep from 35 $^{\circ}$C to 15 $^{\circ}$C and corresponding measured applied heater power showcasing the real-time temperature tracking capabilities of the feedback system.}
\label{fig10}
\end{figure}

\noindent To test the microring locking point, an electronic thermal feedback stabilization circuit is designed and manufactured based on the block diagram of Fig \ref{fig4}. A 2 Gbps modulated optical signal is sent to the input of the double bus microring modulator and the weight driver is adjusted to make the drop gain equal to the through gain. With this configuration, the circuit automatically tunes the microring to the laser wavelength where the average photocurrent of the through and drop ports are equal. The resulting transmission spectrum after a laser wavelength sweep shows a constant optical power for both ports due to the action of the thermal feedback stabilization circuit. When compared with the optical spectrum of the microring without the TFSC, the absence of a resonant peak or notch means that the stabilization circuit is properly working by maintaining equal power at both ports. The measured eye diagrams at resonance and at 250 pm of wavelength detuning with and without thermal feedback are shown in Fig. \ref{fig10}. The equal eye openings at the through and the drop ports show how the thermal feedback stabilizes at the desired operating point. If kept at a fixed temperature, detuning the laser wavelength by 1 GHz (~8 pm) completely changes a microring's resonator transmitted intensity. However, with the proposed stabilization circuit, the optical signals in the through and drop ports remain stable showcasing the ability of the designed circuit to stabilize the microring when there are variations in the laser wavelength. 

To obtain the same wavelength range for the TFSC OFF and TFSC ON curves of Fig. \ref{fig10}a we perform two measurements with the following configurations. For the TFSC ON curve, the laser wavelength is chosen to be 1 nm red-shifted with respect to the microring’s resonance wavelength at steady state ambient temperature, 19 $^{\circ}$C. Then, the TFSC is turned on while applying a laser sweep of 500 pm. Note that the speed of the laser swepp is lower than the thermal constant of the feedback loop. The optical transmitted powers of the through and drop ports are measured using a power meter and the values of the TFSC ON curves are plotted. For the TFSC OFF curve, the TEC temperature is increased to 29 $^{\circ}$C to red shift the optical spectrum approximately 700 pm. A 500 pm sweep is performed while the measured values of the optical transmitted powers of through and drop ports are plotted. The circuit's ability to sustain consistent operation across varying thermal conditions for a fixed input wavelength is demonstrated by performing a temperature sweep using a thermoelectric cooler (TEC). The power applied to the integrated heater is measured and plotted in Fig. \ref{fig10}b. 

\begin{figure}[!ht]
\centering
\includegraphics[width=3.4in]{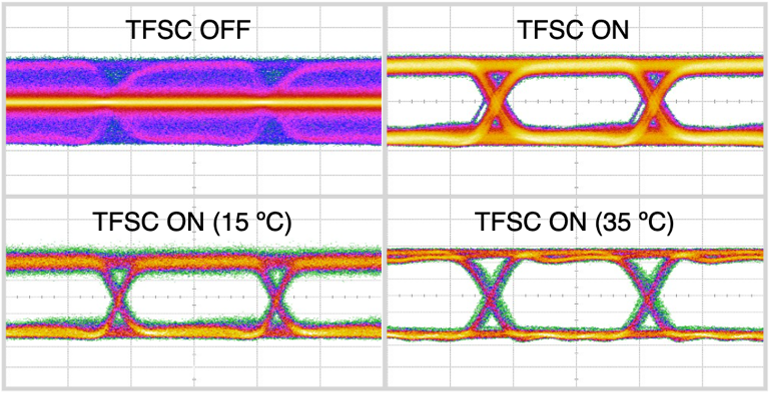}
\caption{Eye diagram at the through port of the MRR after a temperature sweep from 20 $^{\circ}$C to 15 $^{\circ}$C with the TFSC OFF (a) and ON (b) for a 2 Gbps optical input signal. Eye diagrams with the TFSC ON at 15 $^{\circ}$C (c) and 35 $^{\circ}$C (d). For all conditions, a laser of a fixed wavelength is used.}
\label{fig11}
\end{figure}

A measurement of a 2 Gbps transmitted eye diagram in the through port is performed with and without the thermal feedback stabilization circuit connected as shown in Fig. \ref{fig11}. The stability of the tuning and stabilization circuit is evaluated to assess its robustness and reliability. In the experiment, the temperature range for an eye diagram persistence measurement is limited to a change of 5 $^{\circ}$C due to the induced misalignment between the fiber array unit (FAU) and the vertical grating couplers (VGC) given the thermal expansion and contraction of the glue epoxy to fix the DUT to the mechanical stage. It is worth noting that a change in local temperature of less than 1 $^{\circ}$C would end up in the loss of the operating point. To showcase stabilization across a broader temperature range, the stage underwent realignment, and the thermal feedback system is activated at temperatures of 15 $^{\circ}$C and 35 $^{\circ}$C, as illustrated in Fig. \ref{fig11}. When the thermal feedback stabilization circuit is ON, the eye diagrams exhibited minimal variance and distortion across the temperature range, indicative of the circuit's stability and resilience to thermal fluctuations.

\subsection{Weight Function}
\begin{figure}[!ht]
\centering
\includegraphics[width=3.4in]{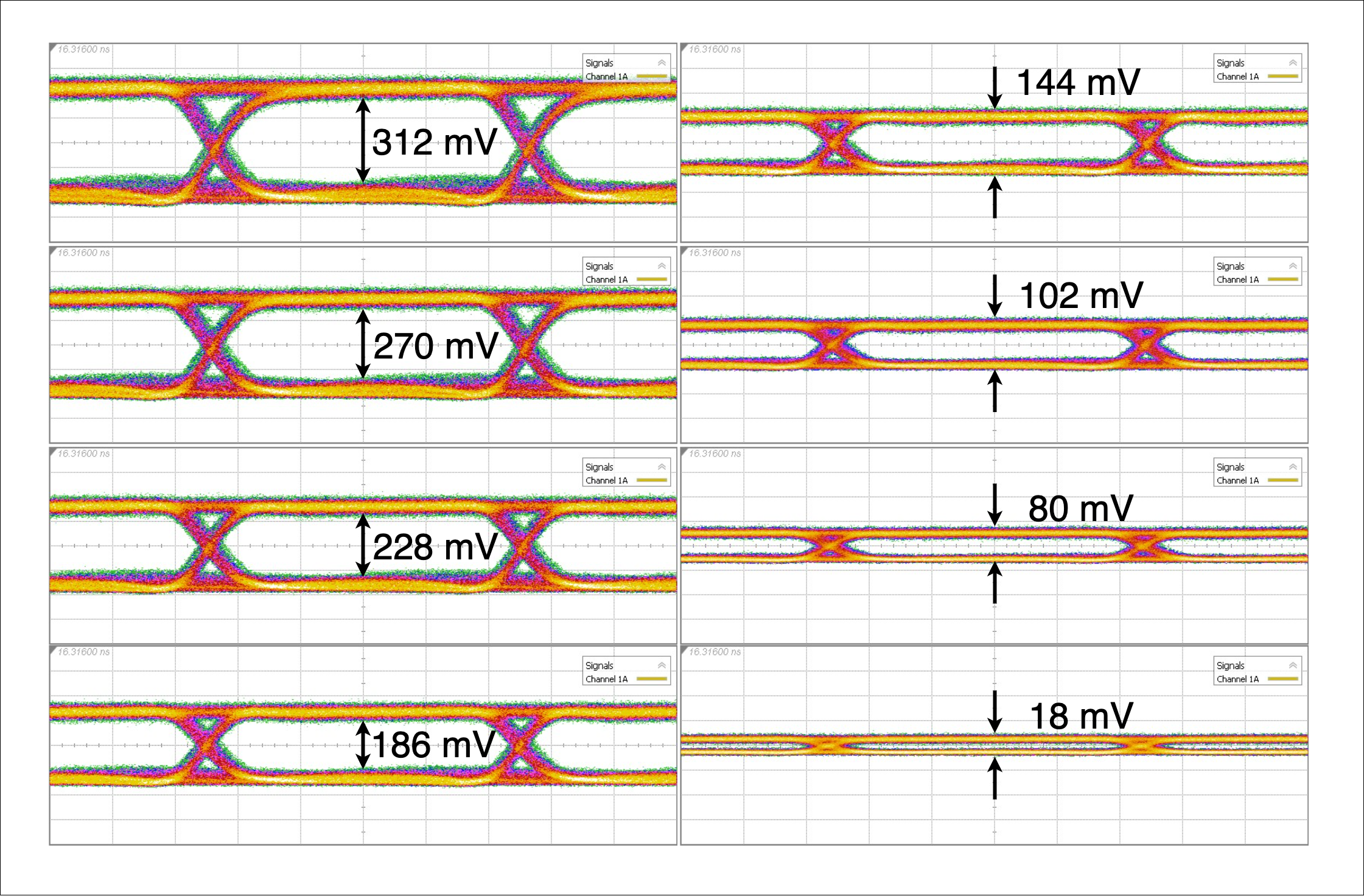}
\caption{Measured eye diagrams at the MRR’s through port for eight different command weights.}
\label{fig12}
\end{figure}

\noindent We evaluate the accuracy and precision of the control by comparing the actual measured weight versus the desired command weight. A set of 16 weight values distributed over a range (0, 1) are defined and applied to the TIA drop gain which translates into a corresponding gain relationship variation between the through and drop TIA’s gain in the thermal feedback loop. As a consequence, this produces a change in the heater’s current leading to the modification of the transmitted weight value that is measured with a power meter while the eye diagram of the 2 Gbps PRBS signal is monitored with the DCA as depicted in Fig. \ref{fig12}. 

\begin{figure}[!ht]
\centering
\includegraphics[width=3.4in]{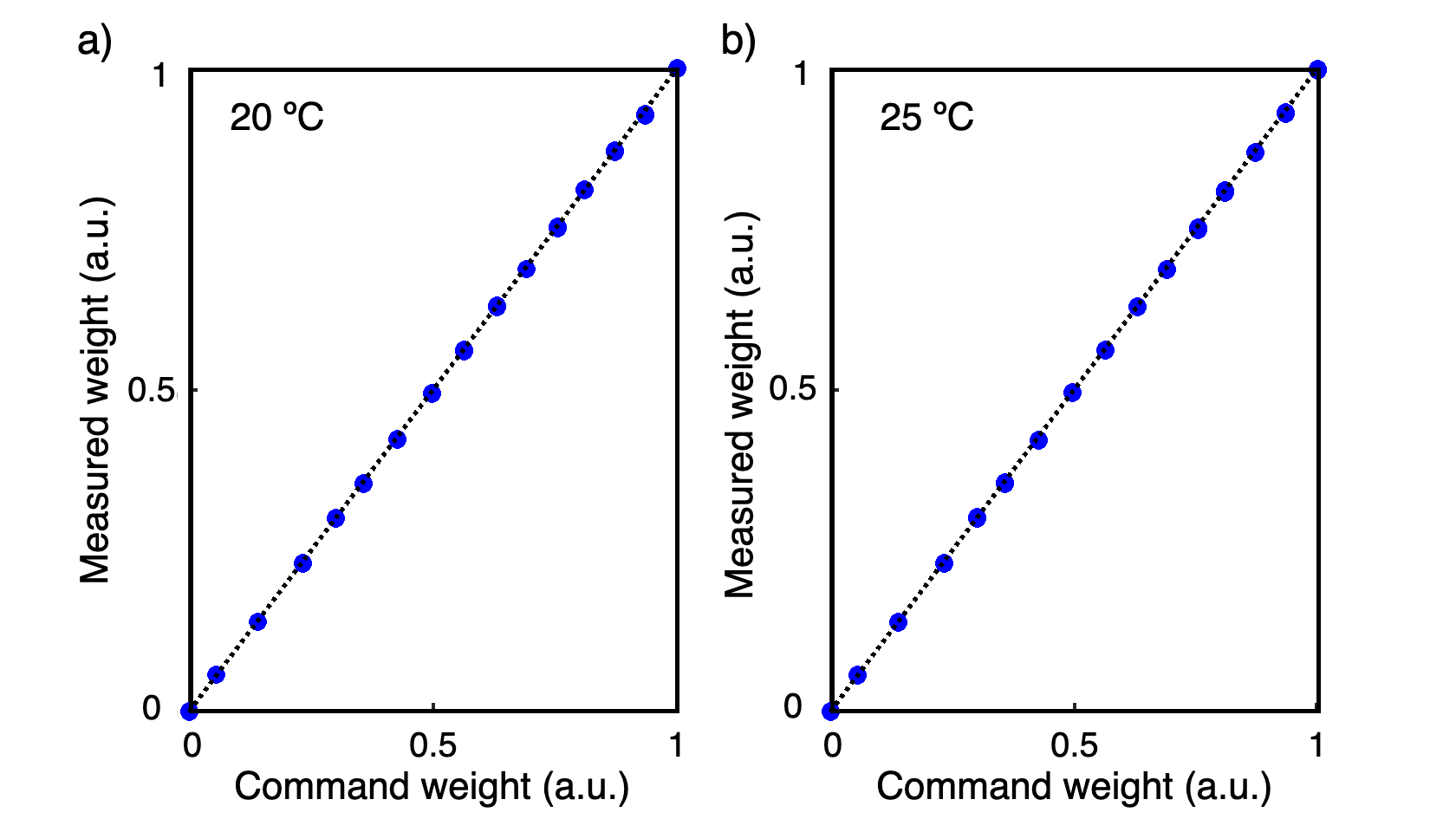}
\caption{Evaluation of the measured weight versus command weight at (a) 20 $^{\circ}$C and (b) 25 $^{\circ}$C compared with the ideal target represented by the black dotted line.}
\label{fig13}
\end{figure}

The performance of the system is quantified in terms of accuracy calculated using the mean absolute error, and precision calculated as the standard deviation over the mean expressed as $\sigma_{accu}=\| \bar\mu - \hat \mu \|$ and $\sigma_{prec}=\sqrt{\langle(\mu - \bar\mu)^2\rangle}$ \cite{shastri}, where $\mu$ is the measured weight, $\bar\mu$ is the mean over the measured weight and $\hat\mu$ is the command weight. Both performance values can be expressed as an effective bit resolution defined as:

\begin{equation}
\label{eq7}
Bit \: Resolution = \log_2(\dfrac{\mu_{max}-\mu_{min}}{\sigma})
\end{equation}
To evaluate the performance, self-calibration and reliability under varying temperatures, the initial calibration and weight mapping are done at 19 $^{\circ}$C while the experimental assessments are conducted at 20 $^{\circ}$C and 25 $^{\circ}$C while the laser wavelength is kept constant. Ten iterations are conducted for each command weight, totaling 160 measurements per temperature. Fig. \ref{fig13} illustrates the measured weights versus command weights, with the ideal curve depicted by the black dotted line and the results of the ten iterations per command weight represented by the blue dots.

\begin{figure}[!ht]
\centering
\includegraphics[width=3.4in]{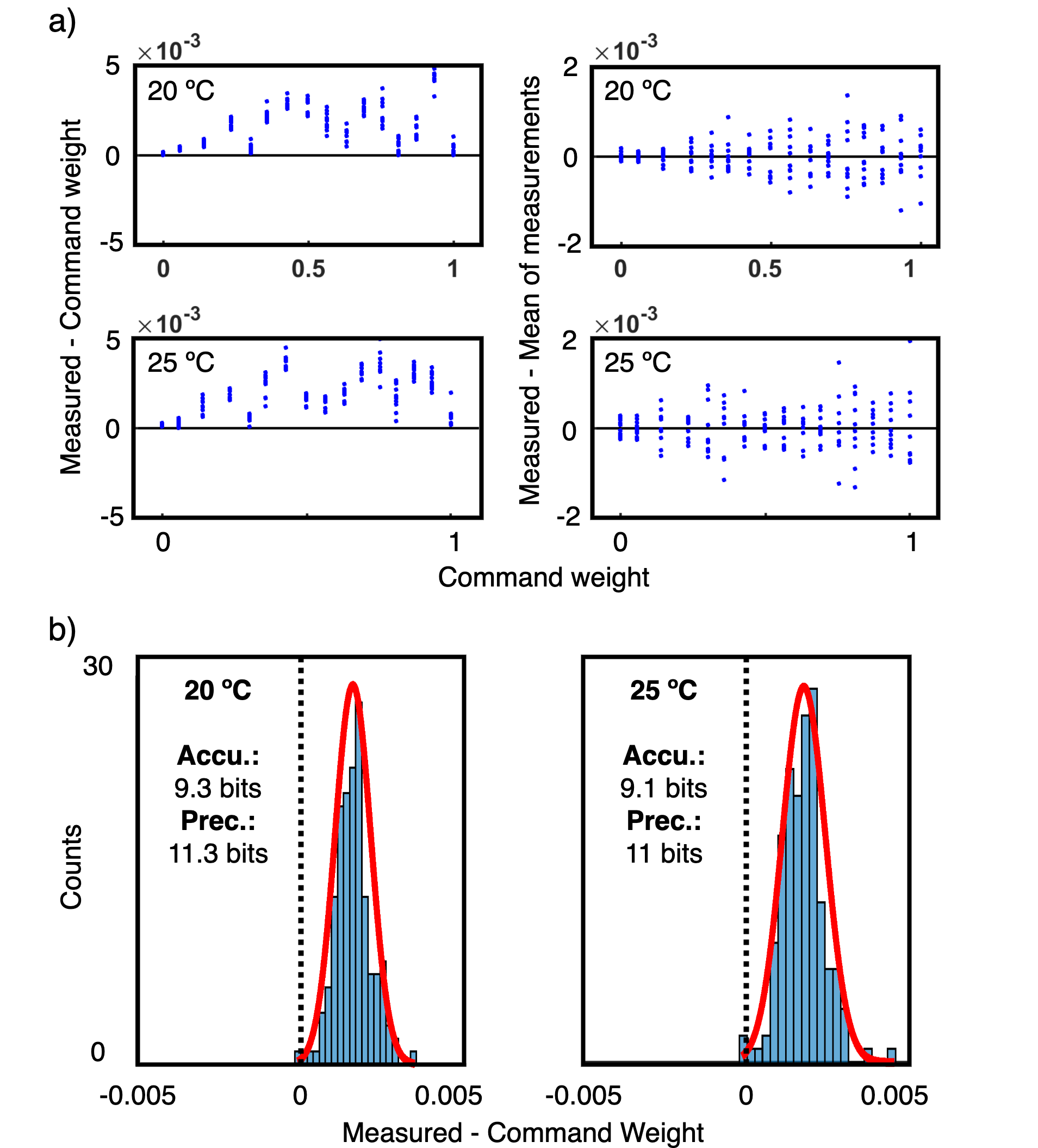}
\caption{a) Scatter representation of accuracy (left) and precision (right) for 20 $^{\circ}$C and 25 $^{\circ}$C. b) probability distribution representation of the experimentally measured accuracy and precision based on the 16 command weights applied at 20 $^{\circ}$C (left) and 25 $^{\circ}$C (right).}
\label{fig14}
\end{figure}

The results revealed mean absolute errors corresponding to resolution values of 9.3 bits and 9.1 bits at 20 $^{\circ}$C and 25 $^{\circ}$C, respectively. Additionally, the precision errors are quantified at 11.3 bits and 11 bits at 20 $^{\circ}$C and 25 $^{\circ}$C, respectively. Scatter points for each measured weight and their histogram representation of accuracy and precision are epicted in Fig. \ref{fig14}. The primary factors contributing to the deterioration of accuracy and precision are the drift of the polarization controllers. This factor results in power coupling loss and a shift in the dynamic range that directly impacted the measured weight values. This effect is visualized in the scatter graph by the increased dispersion observed in the higher command weights compared to the lower command weights.

\begin{figure}[!ht]
\centering
\includegraphics[width=3.4in]{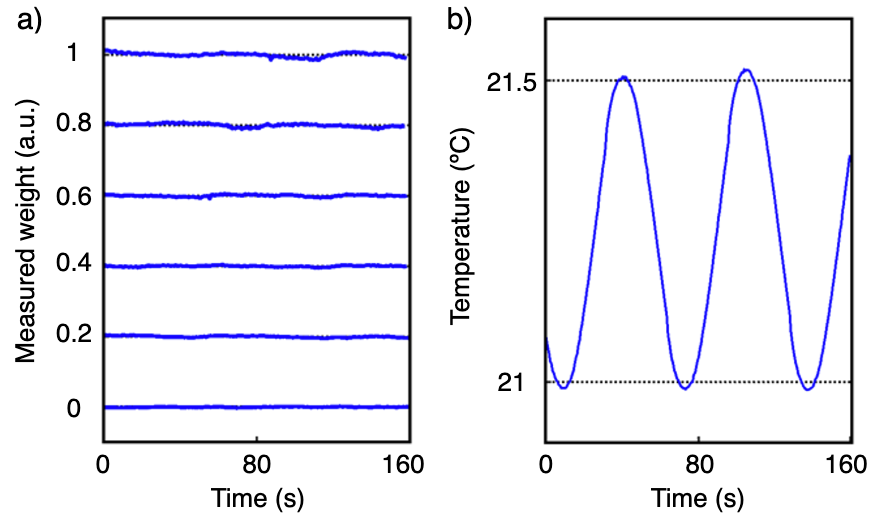}
\caption{a) Real-time stabilization of the six command weights over a duration of 160 seconds, amidst b) sinusoidally varying temperature with 0.5 $^{\circ}$C amplitude.}
\label{fig15}
\end{figure}

Another characterization is conducted by sinusoidally sweeping the set temperature of the thermoelectric cooler (TEC) without realignment of the fiber array and vertical grating couplers to validate the self-calibration potential of the stabilization and control system for real-time temperature variations. The test spanned 160 seconds, with temperature variations limited to 0.5 $^{\circ}$C increments to mitigate fiber to vertical grating coupler misalignments caused by the thermal expansion of the epoxy glue. In this case, six equally spacing command weights from 0 to 1 are applied. Throughout the 160 seconds duration of the test, the system successfully maintained the commanded weight, demonstrating real-time thermal stabilization capabilities under a temperature variation of 0.5 $^{\circ}$C across all six commanded weights. The results highlight the cumulative effect of polarization controller drift, vibrations, and misalignments due to thermal expansion of the epoxy glue, with more pronounced impacts observed as the command weight approaches 1, as shown in Fig. \ref{fig15}.

\section{Discussion}
\noindent Microring tuning and stabilization, alongside accurate, precise, and reconfigurable control, are pivotal for implementing synapse weights in neuromorphic photonics, ensuring learning and plasticity. In our feedback stabilization approach, we propose an all-analog solution utilizing a PID controller to achieve this goal. Table 1 summarizes the comparison between our microring-based weight monitoring design and various previous monitoring approaches. 

\begin{table*}
    \begin{center}
    \caption{Comparison of microring-based weight monitoring schemes}
    \label{tab1}
    \begin{tabular}{c|ccccc}
        Reference & [20] & [30] & [32] & [33] & This work\\ \hline
        Precision & 9 bits & 7.2 bits & 5.4 bits & 7 bits & 11.3 bits \\ \hline
        Immunity to temperature & No & No & No & Yes & Yes\\ 
        fluctuations  &  &  &  &  & \\ \hline
        Self-calibration range of & N/A & N/A & N/A & 0.5 $^{\circ}$C & 6 $^{\circ}$C\\ 
        temperature measured  &  &  &  &  & \\ \hline
        Monitoring scheme & Dithering control &  In-resonator  & In-resonator & Dual-wavelength & Tap coupler \\ 
         &  & photoconductive heater & photoconductive heater & scheme & Self-referenced PID\\ \hline
    \end{tabular}
    \end{center}
\end{table*}

In MRRs, three coupling regimes exist: over-coupling, under-coupling, and critical coupling. Achieving critical coupling maximizes the extinction ratio, leading to an expanded weight function range. This range extension has a positive impact on the resolution of both accuracy and precision, enhancing the system's performance. A degradation in the extinction ratio reduces the bit resolution of the weight function; specifically, every 3-dB reduction in the extinction ratio reduces by two the dynamic range $(\mu_{max}-\mu_{min})$ of the weight function leading to a 1-bit resolution loss.

This tuning, stabilization and control technique is anticipated to facilitate the dense integration of MRR weighted addition by autonomously compensating for thermal crosstalk originating from the integrated heaters of adjacent microrings, as well as for electrical crosstalk stemming from common ground parasitic resistance. Any variations given by these crosstalk phenomena result in changes in the optical intensity of through and drop signals, which are promptly corrected by the feedback stabilization loop. This capability ensures robust performance across densely integrated configurations, enhancing the reliability and scalability of the system without the need for additional calibration efforts.

The maximum range of temperature compensation of the system is determined by several key parameters. The first of which is the thermal tuning efficiency, specified at approximately 100 pm/mW. The second parameter is the maximum current density of the TiW metal trace, limited to approximately 3.3 mA/$\mu$m, defining the upper limit of current flow through the integrated heater, influencing the amount of heat generated. Lastly, the thermal sensitivity, quantified at 70 pm/$^{\circ}$C. In our design, these parameters enable the system to compensate for temperature fluctuations within a range of 60 $^{\circ}$C. In neuromorphic visual systems, where energy efficiency is prioritized over ultra-fast processing speeds, optoelectronic memristors offer a compelling alternative. Their nonvolatile memory, coupled with low power consumption \cite{zhou2,hu} makes them particularly suitable for dense, energy-efficient neuromorphic networks, complementing the fast, parallel processing capabilities of photonic systems.

A TFSC adds complexity, heat dissipation, energy use, and area. However, our all-analog, self-referenced PID controller can be implemented with just a few hundred transistors. Using monolithic electronic-photonic integration (e.g., 45 nm from Global Foundries), the stabilization circuit can be implemented right next to the MRR reducing significantly the metal interconnection length from the control to the MRR, the metal crossings and thus the routing complexity. Additionally, the relatively low speed of the thermal time constant allows to share most of the building blocks of a single high-speed stabilization circuit can to manage multiple microring resonators, mitigating area and energy penalties, thus improving scalability in photonic neuromorphic networks.

Employing Contactless Integrated Photonic Probes (CLIPP) \cite{morichetti} offers a solution to mitigate the power penalty associated with the necessary power splitters for directing signals to the through and drop monitoring photodetectors. This reduction in power loss leads to an augmented dynamic range, consequently enhancing both the accuracy and precision of the system.

Finally, implementing a differential through-drop configuration with independent photodetectors and low-loss microrings could maximize the usable weight range, resulting in an increase in resolution bits. By combining the CLIPP with a differential weight function, a potential increase of two bits in resolution could be achieved.

\section{Conclusion}
\noindent This paper presents a comprehensive design and characterization of an all-analog, self-referenced thermal feedback tuning, stabilization, and weight function control system tailored for neuromorphic computing applications; specifically weighted addition based on MRRs. Through a robust PID-based feedback control an automatic initial tuning and real-time thermal stabilization is achieved. Alongside variable gain TIA’s the system demonstrates the self-calibration of the weight function with an accuracy above 9 bits and a precision above 11 bits for 20 $^{\circ}$C and 25 $^{\circ}$C. The system shows potential to mitigate thermal crosstalk in dense microring systems and CMOS-Photonics monolithic integration. An operating temperature window of 6 $^{\circ}$C for self-calibrated weight function can be achieved ensuring robust performance across varying thermal conditions, dictated by the thermal shifting efficiency and the maximum current density of the integrated heater. Future efforts will focus on adapting the system for differential through – drop for a higher dynamic range as well as multiple simultaneous weight channels, addressing thermal crosstalk challenges in dense microring integration, and exploring known techniques to enhance thermal shifting efficiency.

\end{document}